# Stationary current of a viscous compressed liquid in a cylindrical pipe for the future optical investigations [Стационарное течение вязкой сжимаемой жидкости в цилиндрической трубе]

G.J.Malinowski

[Г.Я.Малиновский]

Most a general view of the equations of movement continuous newton's liquid for lack of volume forces [1]:

$$\rho\cdot\left(\frac{\partial v_i}{\partial t}+v_k\frac{\partial v_i}{\partial x_k}\right)=-\frac{\partial P}{\partial x_i}+\frac{\partial}{\partial x_k}\left\{\eta\cdot\left(\frac{\partial v_i}{\partial x_k}+\frac{\partial v_k}{\partial x_i}-\frac{2}{3}\delta_{ik}\frac{\partial v_l}{\partial x_l}\right)\right\}+\frac{\partial}{\partial x_i}\left(\zeta\cdot\frac{\partial v_l}{\partial x_l}\right) \tag{1}$$

where $\rho$ - density, $\vec{v}$ - speed, $P$ - pressure, $\eta$ and $\zeta$ - the first and the second viscosity (scalar factors). Strain tensor

$$\sigma_{ik}=-p\delta_{ik}+\sigma'_{ik}\ , \tag{2}$$

where $\sigma'_{ik}$ - viscous strain tensor.

In cylindrical co-ordinates $r,\varphi,z$ components viscous strain tensor looks as follows:

$$\left|\begin{array}{lll}
\sigma_{rr}=2\eta\frac{\partial v_r}{\partial r}+\left(\zeta-\frac{2}{3}\eta\right)div\vec{v} & \sigma_{r\varphi}=\eta\left(\frac{1}{r}\frac{\partial v_r}{\partial\varphi}+\frac{\partial v_\varphi}{\partial r}-\frac{v_\varphi}{r}\right) & \sigma_{rz}=\eta\left(\frac{\partial v_z}{\partial r}+\frac{\partial v_r}{\partial z}\right)\\
\sigma_{\varphi r}=\sigma_{r\varphi} & \sigma_{\varphi\varphi}=2\eta\left(\frac{1}{r}\frac{\partial v_\varphi}{\partial\varphi}+\frac{v_r}{r}\right)+\left(\zeta-\frac{2}{3}\eta\right)div\vec{v} & \sigma_{\varphi z}=\eta\left(\frac{\partial v_\varphi}{\partial z}+\frac{1}{r}\frac{\partial v_z}{\partial\varphi}\right)\\
\sigma_{zr}=\sigma_{rz} & \sigma_{z\varphi}=\sigma_{\varphi z} & 2\eta\frac{\partial v_z}{\partial z}+\left(\zeta-\frac{2}{3}\eta\right)div\vec{v}
\end{array}\right| \tag{3}$$

or, attaching the continuity equation

$$\frac{\partial\rho}{\partial t}+\frac{1}{r}\frac{\partial}{\partial r}(r\rho v_r)+\frac{1}{r}\frac{\partial\rho v_\varphi}{\partial\varphi}+\frac{\partial\rho v_z}{\partial z}=0 \tag{4}$$

$$\rho\frac{\partial v_r}{\partial t}+\rho\cdot\left(v_r\frac{\partial v_r}{\partial r}+\frac{v_\varphi}{r}\frac{\partial v_r}{\partial\varphi}+v_z\frac{\partial v_r}{\partial z}-\frac{v_\varphi^2}{r}\right)=-\frac{\partial P}{\partial r}+\eta\cdot\left[\frac{1}{r}\frac{\partial}{\partial r}\left(r\frac{\partial v_r}{\partial r}\right)+\frac{1}{r^2}\frac{\partial^2 v_r}{\partial\varphi^2}+\frac{\partial^2 v_r}{\partial z^2}-\frac{2}{r^2}\frac{\partial v_\varphi}{\partial\varphi}-\frac{v_r}{r^2}\right]+$$

$$+\frac{\partial\eta}{\partial r}\cdot\left[2\frac{\partial v_r}{\partial r}-\frac{2}{3}\frac{1}{r}\frac{\partial}{\partial r}(rv_r)-\frac{2}{3}\frac{1}{r}\frac{\partial v_\varphi}{\partial\varphi}-\frac{2}{3}\frac{\partial v_z}{\partial z}\right]+\frac{1}{r}\frac{\partial\eta}{\partial\varphi}\cdot\left(\frac{\partial v_\varphi}{\partial r}-\frac{v_\varphi}{r}+\frac{1}{r}\frac{\partial v_r}{\partial\varphi}\right)+\frac{\partial\eta}{\partial z}\cdot\left(\frac{\partial v_r}{\partial z}+\frac{\partial v_z}{\partial r}\right)+$$

$$+\frac{\partial\zeta}{\partial r}\cdot\left[\frac{1}{r}\frac{\partial}{\partial r}(rv_r)+\frac{1}{r}\frac{\partial v_\varphi}{\partial\varphi}+\frac{\partial v_z}{\partial z}\right]+\left(\zeta+\frac{1}{3}\eta\right)\cdot\frac{\partial}{\partial r}\left(\frac{1}{r}\frac{\partial}{\partial r}(rv_r)+\frac{1}{r}\frac{\partial v_\varphi}{\partial\varphi}+\frac{\partial v_z}{\partial z}\right) \tag{5}$$

$$\rho\frac{\partial v_\varphi}{\partial t}+\rho\cdot\left(v_r\frac{\partial v_\varphi}{\partial r}+\frac{v_\varphi}{r}\frac{\partial v_\varphi}{\partial\varphi}+v_z\frac{\partial v_\varphi}{\partial z}-\frac{v_r v_\varphi}{r}\right)=-\frac{1}{r}\frac{\partial P}{\partial\varphi}+\eta\cdot\left[\frac{1}{r}\frac{\partial}{\partial r}(r\frac{\partial v_\varphi}{\partial r})+\frac{1}{r^2}\frac{\partial^2 v_\varphi}{\partial\varphi^2}+\frac{\partial^2 v_\varphi}{\partial z^2}+\frac{2}{r^2}\frac{\partial v_r}{\partial\varphi}-\frac{v_\varphi}{r^2}\right]+$$

$$+\frac{\partial\eta}{\partial r}\cdot\left(\frac{\partial v_\varphi}{\partial r}-\frac{v_\varphi}{r}+\frac{1}{r}\frac{\partial v_r}{\partial\varphi}\right)+\frac{1}{r}\frac{\partial\eta}{\partial\varphi}\cdot\left[\frac{2}{r}v_r+\frac{2}{r}\frac{\partial v_\varphi}{\partial\varphi}-\frac{2}{3}\frac{1}{r}\frac{\partial}{\partial r}\left(rv_r\right)-\frac{2}{3}\frac{1}{r}\frac{\partial v_\varphi}{\partial\varphi}-\frac{2}{3}\frac{\partial v_z}{\partial z}\right]+\frac{\partial\eta}{\partial z}\cdot\left(\frac{\partial v_\varphi}{\partial z}+\frac{1}{r}\frac{\partial v_z}{\partial\varphi}\right)+$$

$$+\frac{1}{r}\frac{\partial\zeta}{\partial\varphi}\cdot\left[\frac{1}{r}\frac{\partial}{\partial r}(rv_r)+\frac{1}{r}\frac{\partial v_\varphi}{\partial\varphi}+\frac{\partial v_z}{\partial z}\right]+(\zeta+\frac{1}{3}\eta)\cdot\frac{1}{r}\frac{\partial}{\partial\varphi}\left(\frac{1}{r}\frac{\partial}{\partial r}(rv_r)+\frac{1}{r}\frac{\partial v_\varphi}{\partial\varphi}+\frac{\partial v_z}{\partial z}\right) \tag{6}$$

$$\rho\frac{\partial v_z}{\partial t}+\rho\cdot\left(v_r\frac{\partial v_z}{\partial r}+\frac{v_\varphi}{r}\frac{\partial v_z}{\partial\varphi}+v_z\frac{\partial v_z}{\partial z}\right)=-\frac{\partial P}{\partial z}+\eta\cdot\left[\frac{1}{r}\frac{\partial}{\partial r}(r\frac{\partial v_z}{\partial r})+\frac{1}{r^2}\frac{\partial^2 v_z}{\partial\varphi^2}+\frac{\partial^2 v_z}{\partial z^2}\right]+$$

$$+\frac{\partial\eta}{\partial r}\cdot\left(\frac{\partial v_r}{\partial z}+\frac{\partial v_z}{\partial r}\right)+\frac{1}{r}\frac{\partial\eta}{\partial\varphi}\cdot\left(\frac{\partial v_\varphi}{\partial z}+\frac{1}{r}\frac{\partial v_z}{\partial\varphi}\right)+\frac{\partial\eta}{\partial z}\cdot\left[2\frac{\partial v_z}{\partial z}-\frac{2}{3}\frac{1}{r}\frac{\partial}{\partial r}\left(rv_r\right)-\frac{2}{3}\frac{1}{r}\frac{\partial v_\varphi}{\partial\varphi}-\frac{2}{3}\frac{\partial v_z}{\partial z}\right]+$$

$$+\frac{\partial\zeta}{\partial z}\cdot\left[\frac{1}{r}\frac{\partial}{\partial r}(rv_r)+\frac{1}{r}\frac{\partial v_\varphi}{\partial\varphi}+\frac{\partial v_z}{\partial z}\right]+(\zeta+\frac{1}{3}\eta)\cdot\frac{\partial}{\partial z}\left(\frac{1}{r}\frac{\partial}{\partial r}(rv_r)+\frac{1}{r}\frac{\partial v_\varphi}{\partial\varphi}+\frac{\partial v_z}{\partial z}\right) \tag{7}$$

Let's consider the approach following:

1. $\frac{\partial}{\partial t}\equiv 0$ (stationarity)
2. $\frac{\partial P}{\partial r}=0$ and $v_r=0$ (absence of radial movement)
3. $\frac{\partial}{\partial\varphi}\equiv 0$ (angular symmetry)

Therefore

$$\frac{\partial\rho v_z}{\partial z}=0 \tag{8}$$

$$-\rho\frac{v_\varphi^2}{r}=-\frac{2}{3}\frac{\partial\eta}{\partial r}\frac{\partial v_z}{\partial z}+\frac{\partial\eta}{\partial z}\frac{\partial v_z}{\partial r}+\frac{\partial\zeta}{\partial r}\frac{\partial v_z}{\partial z}+(\zeta+\frac{1}{3}\eta)\cdot\frac{\partial^2 v_z}{\partial r\partial z} \tag{9}$$

$$\rho v_z\frac{\partial v_\varphi}{\partial z}=\eta\cdot\left[\frac{1}{r}\frac{\partial}{\partial r}(r\frac{\partial v_\varphi}{\partial r})+\frac{\partial^2 v_\varphi}{\partial z^2}-\frac{v_\varphi}{r^2}\right]+\frac{\partial\eta}{\partial r}\cdot\left(\frac{\partial v_\varphi}{\partial r}-\frac{v_\varphi}{r}\right)+\frac{\partial\eta}{\partial z}\frac{\partial v_\varphi}{\partial z} \tag{10}$$

$$\rho v_z\frac{\partial v_z}{\partial z}=-\frac{dP}{dz}+\eta\cdot\frac{1}{r}\frac{\partial}{\partial r}(r\frac{\partial v_z}{\partial r})+\frac{\partial\eta}{\partial r}\frac{\partial v_z}{\partial r}+\frac{4}{3}\frac{\partial\eta}{\partial z}\frac{\partial v_z}{\partial z}+\frac{\partial\zeta}{\partial z}\frac{\partial v_z}{\partial z}+(\zeta+\frac{4}{3}\eta)\cdot\frac{\partial^2 v_z}{\partial z^2} \tag{11}$$

From (8) it is had:

$$\rho v_z = f(r) \tag{12}$$

Let's search therefore for the decision a method of division of variables[2] :

$$v_\varphi(r,z) = R_\varphi(r) Z_\varphi(z) \tag{13}$$

$$v_z(r,z) = R(r)Z(z) \tag{14}$$

$$\rho(r,z) = \rho_r(r)\rho_z(z) \tag{15}$$

$$\eta(r,z) = \eta_r(r)\eta_z(z) \tag{16}$$

$$\zeta(r,z) = \zeta_r(r)\zeta_z(z) \tag{17}$$

From (12), having cleaned dimensions in radial variables:

$$\rho_z Z = \frac{f(r)}{\rho_r R} = const = 1 \tag{18}$$

or

$$v_z = \frac{R}{\rho_z} = R(r)y(z), \tag{19}$$

where for $y \equiv \frac{1}{\rho_z}$ have designated return z-density.

$$-\rho_r \frac{R_\varphi^2 Z_\varphi^2}{ry} = -\frac{2}{3}\eta_r^{'}\eta_z R y^{'} + \eta_r \eta_z^{'} R^{'} y + \zeta_r^{'}\zeta_z R y^{'} + (\zeta_r \zeta_z + \frac{1}{3}\eta_r\eta_z)\cdot R^{'} y^{'} \tag{20}$$

$$\rho_r R R_\varphi Z_\varphi^{'} = \eta_r \eta_z \cdot \left( \frac{1}{r}(rR_\varphi^{'})^{'} Z_\varphi + R_\varphi Z_\varphi^{''} - \frac{R_\varphi Z_\varphi}{r^2} \right) + \eta_r^{'}\eta_z \cdot \left( R_\varphi^{'} - \frac{R_\varphi}{r} \right) \cdot Z_\varphi + \eta_r \eta_z^{'} R_\varphi Z_\varphi^{'} \tag{21}$$

$$\rho_r R^2 y^{'} = -P^{'} + \eta_r \eta_z \cdot \frac{1}{r}(rR^{'})^{'} \cdot y + \eta_r^{'}\eta_z R^{'} y + \frac{4}{3}\eta_r \eta_z^{'} R y^{'} + \zeta_r \zeta_z^{'} R y^{'} + (\zeta_r \zeta_z + \frac{4}{3}\eta_r \eta_z)\cdot R y^{''} \tag{22}$$

Division of variables exists at additional approach of identical dependence from r of the both viscosities in addition to three available:

4. $\zeta_r(r) = \alpha \eta_r(r)$ (23)

$$-\rho_r \frac{R_\varphi^2}{r} \cdot \frac{\eta_z Z_\varphi^2}{y y^{'}} = -\eta_r^{'} R \cdot \eta_z^{2} + \eta_r R^{'} \cdot \eta_z \eta_z^{'} \frac{y}{y^{'}} + (\eta_r R)^{'} \cdot \eta_z (\alpha \zeta_z + \frac{1}{3}\eta_z) \tag{24}$$

$$\rho_r R R_\varphi \cdot \frac{Z_\varphi^{'}}{\eta_z Z_\varphi} = \eta_r \left( \frac{1}{r}(rR_\varphi^{'})^{'} - \frac{R_\varphi}{r^2} \right) + \eta_r^{'} \left( R_\varphi^{'} - \frac{R_\varphi}{r} \right) + \eta_r R_\varphi \cdot \left( \frac{Z_\varphi^{''}}{Z_\varphi} + \frac{\eta_z^{'}}{\eta_z}\frac{Z_\varphi^{'}}{Z_\varphi} \right) \tag{25}$$

$$\rho_r R^2 \cdot \frac{y^{'}}{\eta_z y} = -\frac{P^{'}}{\eta_z y} + \eta_r \frac{1}{r}(rR^{'})^{'} + \eta_r^{'} R^{'} + \eta_r R \cdot \frac{\left( (\alpha\zeta_z + \frac{4}{3}\eta_z) y^{'} \right)^{'}}{\eta_z y} \tag{26}$$

Let's search for the decision for (24) under a condition:

5. $\eta_r^{'} R = \beta \eta_r R^{'}$ (27)

** we will notice that instead of 5. there can be and another, stronger, approach when we neglect the dependence of the viscosities from radius:

5'. $\eta_r^{'} = \zeta_r^{'} = 0$ (28)

Systems of the ordinary differential equations on z looks like for 5.:

$$\begin{cases} \dfrac{y'}{\eta_z y} = -C_1 \\ \dfrac{\left((\alpha\zeta_z + \frac{4}{3}\eta_z)y'\right)'}{\eta_z y} = C_2 \\ -\dfrac{P'}{\eta_z y} = C_3 \\ -\dfrac{1}{C_1}\eta_z' + \eta_z(\alpha\zeta_z + \frac{1}{3}\eta_z) + \beta\eta_z(\alpha\zeta_z - \frac{2}{3}\eta_z) = C_4 \end{cases} \qquad (29)$$

$$\begin{cases} \dfrac{\eta_z Z_\varphi^2}{yy'} = -C_5 \\ \dfrac{Z_\varphi'}{\eta_z Z_\varphi} = -C_6 \\ \dfrac{Z_\varphi''}{Z_\varphi} + \dfrac{\eta_z'}{\eta_z}\dfrac{Z_\varphi'}{Z_\varphi} = C_7 \end{cases} \qquad (30)$$

Accordingly on r:

$$\begin{cases} \eta_r' R = \beta\eta_r R' \\ C_5\rho_r \dfrac{R_\varphi^2}{r} = C_4\eta_r R' \\ -C_6\rho_r R R_\varphi = \eta_r\left(\dfrac{1}{r}(rR_\varphi')' - \dfrac{R_\varphi}{r^2}\right) + \eta_r'\left(R_\varphi' - \dfrac{R_\varphi}{r}\right) + C_7\eta_r R_\varphi \\ \eta_r \dfrac{1}{r}(rR')' + \eta_r' R' + C_1\rho_r R^2 + C_2\eta_r R + C_3 = 0 \end{cases} \qquad (31)$$

The system (29)is closed on four unknown variables $\eta_z(z), y(z), P(z)$ and $\zeta_z(z)$. Consecutive expression of one variable through other leads to the ordinary nonlinear differential equation of the second order for $\eta_z(z)$:

$$\eta_z'' + C_1(1+4\beta)\eta_z\eta_z' - C_1^2(1+2\beta)\eta_z^3 + \left(C_2(1+\beta) - C_4C_1^2\right)\eta_z = 0 \qquad (32)$$

After replacing $w(\xi) = \eta_z'$ also $\xi = -C_1(1+4\beta)/2\cdot\eta_z^2 + \dfrac{[C_2(1+\beta) - C_4C_1^2](1+4\beta)}{2C_1(1+2\beta)}$ we receive Abel's equation

$$ww' - w = a\xi, \qquad (33)$$

where

$$a = \frac{2(1+2\beta)}{(1+4\beta)^2} \qquad (34)$$

*Owing to similarity of "weight" w ' and 1 we will search for the decision* (33) *through the linear functions:*

$$(w+\gamma_1\xi)^{\gamma_2}\cdot(w-\gamma_3\xi)^{\gamma_4} = C$$

*Or differentiating,*

$$\gamma_2(w+\gamma_1\xi)^{\gamma_2-1}\cdot(w-\gamma_3\xi)^{\gamma_4}\cdot(w'+\gamma_1) + \gamma_4(w+\gamma_1\xi)^{\gamma_2}\cdot(w-\gamma_3\xi)^{\gamma_4-1}\cdot(w'-\gamma_3) = 0$$

$$\gamma_2(w-\gamma_3\xi)\cdot(w'+\gamma_1) + \gamma_4(w+\gamma_1\xi)\cdot(w'-\gamma_3) = 0$$

$$\gamma_2 ww' + \gamma_2\gamma_1 w - \gamma_2\gamma_3\xi w' - \gamma_2\gamma_3\gamma_1\xi + \gamma_4 ww' - \gamma_4\gamma_3 w + \gamma_4\gamma_1\xi w' - \gamma_4\gamma_1\gamma_3\xi = 0$$

*Whence*

$\gamma_1\gamma_4 = \gamma_2\gamma_3$

$\gamma_2 + \gamma_4 = \gamma_3\gamma_4 - \gamma_1\gamma_2$

$\gamma_3 - \gamma_1 = 1$

*or*

$\gamma_1 = \gamma_2$ *And* $\gamma_3 = \gamma_4$

*Hence, the common decision of the equation of Abel* (33)

$ww' - w = a\xi$

*Equally*

$$[w+(\gamma-1)\xi]^{\gamma-1}\cdot(w-\gamma\xi)^{\gamma} = C, \quad (35)$$

*where*

$$a = \gamma(\gamma-1) \quad (36)$$

For ours (34) we have a quadratic equation

$$\gamma(\gamma-1) = a = \frac{2(1+2\beta)}{(1+4\beta)^2} \quad (37)$$

or

$$\gamma_1 = \frac{2(1+2\beta)}{(1+4\beta)} \text{ and } \gamma_2 = -\frac{1}{(1+4\beta)} \quad (38)$$

Definitively, implicit equation for definition of viscosity $\eta_z(z)$ from (32)

$$\left[\eta_z' - \frac{C_1}{2}\cdot\left(\eta_z^2 - \frac{[C_2(1+\beta)-C_4C_1^2]}{C_1^2(1+2\beta)}\right)\right]^{\frac{1}{(1+4\beta)}} \cdot \left[\eta_z' + C_1(1+2\beta)\cdot\left(\eta_z^2 - \frac{[C_2(1+\beta)-C_4C_1^2]}{C_1^2(1+2\beta)}\right)\right]^{\frac{2(1+2\beta)}{(1+4\beta)}} = C \quad (39)$$

for the first root or it is symmetric for the second root

$$\left[\eta_z' - \frac{C_1}{2}\cdot\left(\eta_z^2 - \frac{[C_2(1+\beta)-C_4C_1^2]}{C_1^2(1+2\beta)}\right)\right]^{-\frac{2(1+2\beta)}{(1+4\beta)}} \cdot \left[\eta_z' + C_1(1+2\beta)\cdot\left(\eta_z^2 - \frac{[C_2(1+\beta)-C_4C_1^2]}{C_1^2(1+2\beta)}\right)\right]^{-\frac{1}{(1+4\beta)}} = C \quad (40)$$

So, the system (29) has non-trivial decision.

Now we will consider system (30) for z-components $Z_\varphi$ of speed $v_\varphi$. I will remind that we used at its conclusion approach 3. of the angular symmetry. For its performance the constancy of angular speed is necessary at least

$$R_\varphi = \omega\cdot r \quad (41)$$

or in general absence of rotation

$$v_\varphi \equiv R_\varphi Z_\varphi = 0 \quad (42)$$

Depending on what zero "is more senior" - on r or on z, the system (30) or will have non-trivial decision, or is trivial $C_5 = C_6 = C_7 = 0$

With such reservation also we will consider system(30).

From the first equation(29):

$$\eta_z = -\frac{1}{C_1}\frac{y'}{y} \quad (43)$$

$$\begin{cases} Z_\varphi = \sqrt{C_1C_5} \cdot y \\ C_6 = C_1 \\ \frac{y''}{y} + \frac{\eta_z'}{\eta_z}\frac{y'}{y} = C_7 \end{cases} \tag{44}$$

i.e. z-components of speeds $v_\varphi$ also $v_z$ are identical, and for viscosity $\eta_z$ the simple condition is imposed

$$\eta_z' - \frac{C_1}{2} \cdot \left(\eta^2 - \frac{C_7}{C_1^2}\right) = 0 \tag{45}$$

It corresponds to the equation (39) for $C = 0$ and $C_7 = \frac{[C_2(1+\beta) - C_4C_1^2]}{(1+2\beta)}$, (46)

For a r-component of system (31) with the reservation(41):

$$\begin{cases} \eta_r' R = \beta\eta_r R' \\ C_5\rho_r\omega^2 r = C_4(\omega) \cdot R' \\ -C_1\rho_r R = C_7\eta_r \\ \eta_r \frac{1}{r}(rR')' + \eta_r' R' + C_1\rho_r R^2 + C_2\eta_r R + C_3 = 0 \end{cases} \tag{47}$$

or

$$\frac{1}{r}(rR')' + \beta\frac{R'^2}{R} + \frac{\beta C_2 + C_4C_1^2}{1+2\beta} \cdot R + C_3 \cdot R^{-\beta} = 0 \ , \tag{48}$$

with boundary conditions

$$R(0) = 1 \text{ and } R'(0) = 0 \tag{49}$$

has the decision after which definition are both radial viscosity and density:

$$\eta_r = aR^\beta \tag{50}$$

$$\zeta_r = \alpha\eta_r \tag{51}$$

$$\rho_r = -\frac{C_7}{C_1}\frac{\eta_r}{R} \tag{52}$$

Let's notice that in the absence of rotation and at $\beta \to 0$, i.e. at independence $\eta$ and $\zeta$ from $r$, - the equation from (48) naturally passes during *Poiseuille*:

$$\frac{1}{r}(rR')' + C_3 = 0 \tag{53}$$

with the decision

$$R = \frac{C_3 r_0^2}{4}\left(1 - r^2\right) \tag{54}$$

Let's return now to z-components. We will substitute C=0 in (39) or (40), thus we will consider from last equation (29) the entry condition

$$\eta_0' = C_1\eta_0 L \cdot \left(\frac{1}{3} + (1+\beta)\alpha\frac{\zeta_0}{\eta_0} - \frac{2\beta}{3} - \frac{C_4}{\eta_0^2}\right) \tag{55}$$

Then or

$$C_2 = \left[\frac{3+4\beta}{1+\beta}\frac{C_4}{\eta_0^2} + \frac{(1+4\beta)(1+2\beta)}{3(1+\beta)} - 2\alpha\frac{\zeta_0}{\eta_0}(1+2\beta)\right] \cdot C_1^2\eta_0^2 \tag{56}$$

or

$$C_2 = \left( \frac{4}{3} + \alpha \frac{\zeta_0}{\eta_0} \right) \cdot C_1^2 \eta_0^2 \quad \text{and} \quad \beta = -\frac{3}{4} \tag{57}$$

The equation (45) solves in simple hyperbolic or trigonometrical functions for resulted on initial at $z = 0$ value of sizes like $x = \frac{X}{X_0}$. The top sign answers a positive sign $C_1$, bottom – negative:

$$\varepsilon^2 = 2\frac{C_4}{\eta_0^2} + \frac{1}{3}(1+4\beta) - 2\alpha \frac{\zeta_0}{\eta_0}(1+\beta) \tag{58}$$

$$\delta = \frac{1}{2}|C_1| \eta_0 L \varepsilon \tag{59}$$

a) $\varepsilon > 1$ (60)

$$\eta_z = \frac{th(\delta(c \mp z))}{th(\delta c)} \tag{61}$$

$$\rho_z \equiv \frac{1}{y} = \frac{ch^2(\delta c)}{ch^2(\delta(c \mp z))} \tag{62}$$

$$\zeta_z = \frac{\frac{C_4 C_1^2 L^2}{4\delta^2} cth(\delta(c \mp z)) - \frac{1}{sh(2\delta(c \mp z))} - \frac{1-2\beta}{3} th(\delta(c \mp z))}{\frac{C_4 C_1^2 L^2}{4\delta^2} cth(\delta c) - \frac{1}{sh(2\delta c)} - \frac{1-2\beta}{3} th(\delta c)}$$

(63)

$$P = \frac{ch^2(\delta(c \mp z)) - ch^2(\delta(c \mp 1))}{ch^2(\delta c) - ch^2(\delta(c \mp 1))} \tag{64}$$

where

$$c = \frac{1}{\delta} Arth\left(\frac{1}{\varepsilon}\right) \tag{65}$$

b) $0 < \varepsilon < 1$ (66)

$$\eta_z = \frac{cth(\delta(c \mp z))}{cth(\delta c)} \tag{67}$$

$$\rho_z \equiv \frac{1}{y} = \frac{sh^2(\delta c)}{sh^2(\delta(c \mp z))} \tag{68}$$

$$\zeta_z = \frac{\frac{C_4 C_1^2 L^2}{4\delta^2} th(\delta(c \mp z)) + \frac{1}{sh(2\delta(c \mp z))} - \frac{1-2\beta}{3} cth(\delta(c \mp z))}{\frac{C_4 C_1^2 L^2}{4\delta^2} th(\delta c) + \frac{1}{sh(2\delta c)} - \frac{1-2\beta}{3} cth(\delta c)}$$

(69)

$$P = \frac{sh^2(\delta(c \mp z)) - sh^2(\delta(c \mp 1))}{sh^2(\delta c) - sh^2(\delta(c \mp 1))} \tag{70}$$

where

$$c = \frac{1}{\delta} Arth(\varepsilon) \tag{71}$$

c) $$\varepsilon^2 = 2\alpha\frac{\zeta_0}{\eta_0}(1+\beta) - \frac{1}{3}(1+4\beta) - 2\frac{C_4}{\eta_0^2} \quad (72)$$

$$\eta_z = \frac{tg\left(\delta(c \pm z)\right)}{tg\left(\delta c\right)} \quad (73)$$

$$\rho_z \equiv \frac{1}{y} = \frac{\cos^2\left(\delta c\right)}{\cos^2\left(\delta(c \pm z)\right)} \quad (74)$$

$$\zeta_z = \frac{\frac{C_4 C_1^2 L^2}{4\delta^2} ctg\left(\delta(c \pm z)\right) + \frac{1}{\sin\left(2\delta(c \pm z)\right)} - \frac{1-2\beta}{3} tg\left(\delta(c \pm z)\right)}{\frac{C_4 C_1^2 L^2}{4\delta^2} ctg\left(\delta c\right) + \frac{1}{\sin\left(2\delta c\right)} - \frac{1-2\beta}{3} tg\left(\delta c\right)} \quad (75)$$

$$P = \frac{\cos^2\left(\delta(c \pm z)\right) - \cos^2\left(\delta(c \pm 1)\right)}{\cos^2\left(\delta c\right) - \cos^2\left(\delta(c \pm 1)\right)} \quad (76)$$

where

$$c = \frac{1}{\delta} arctg\left(\frac{1}{\varepsilon}\right) \quad (77)$$

Depending on a sign $C_1$ the density along a pipe either grows, or falls:

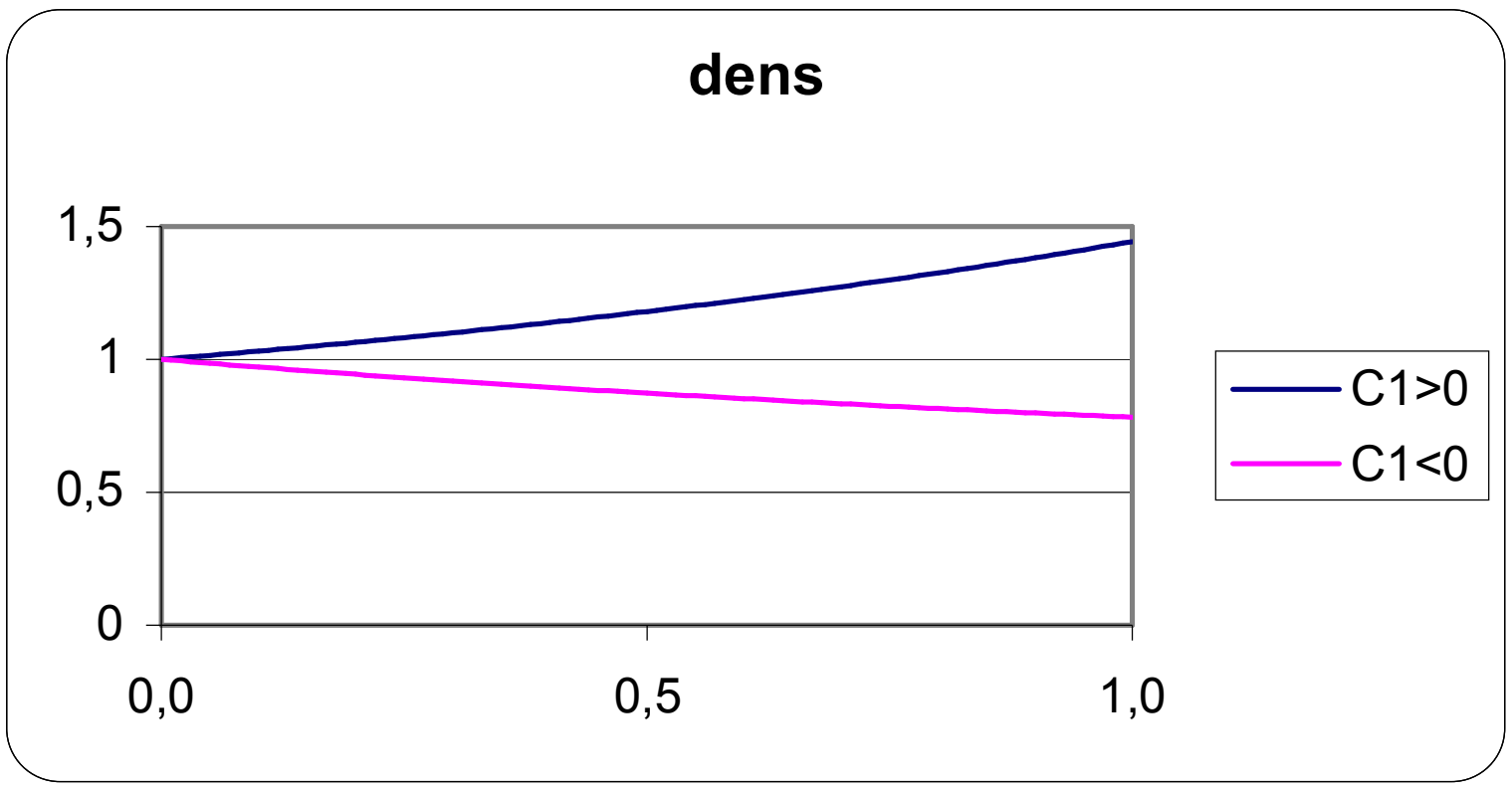

And pressure has or a concave kind

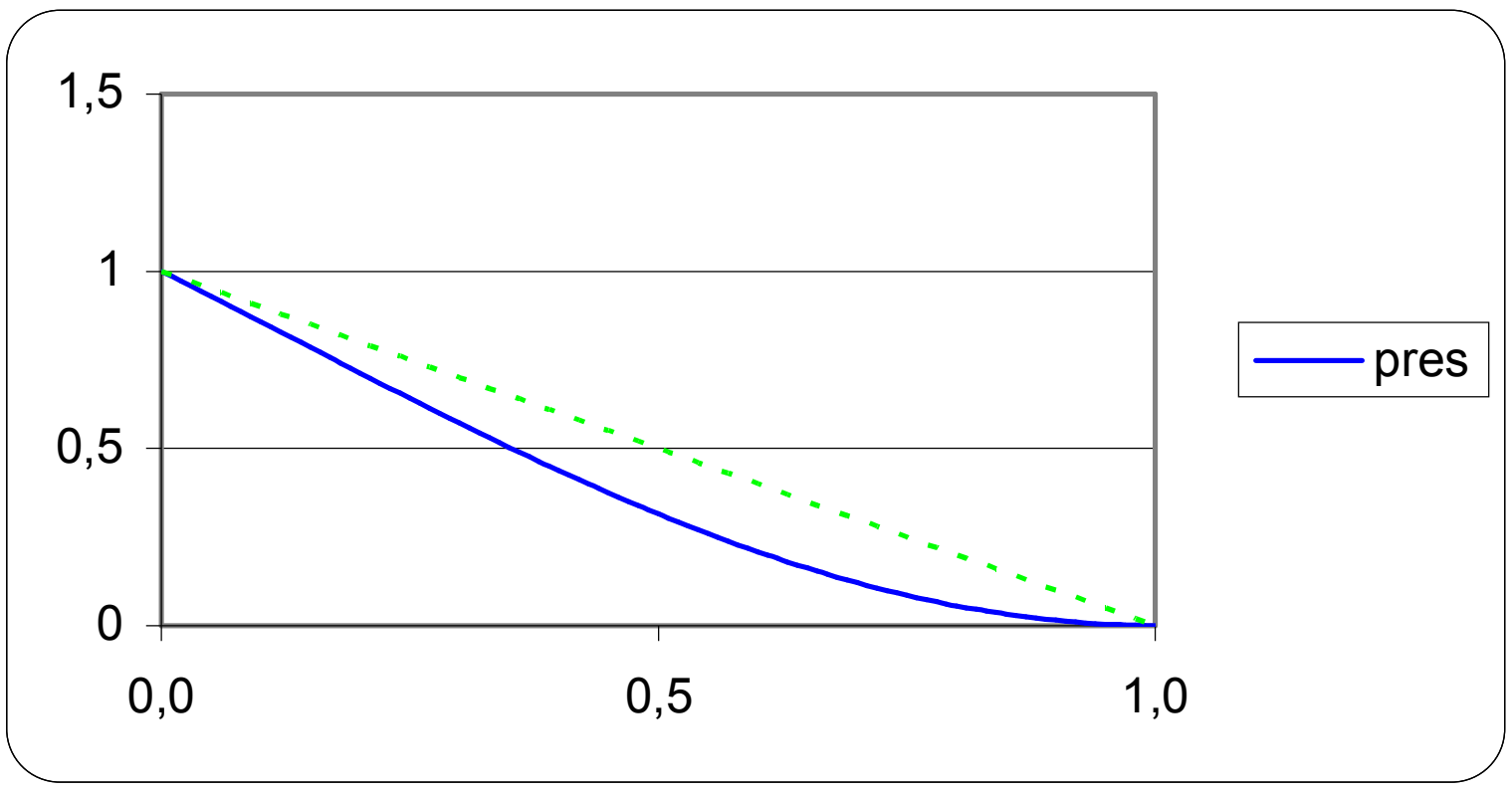

or convex one

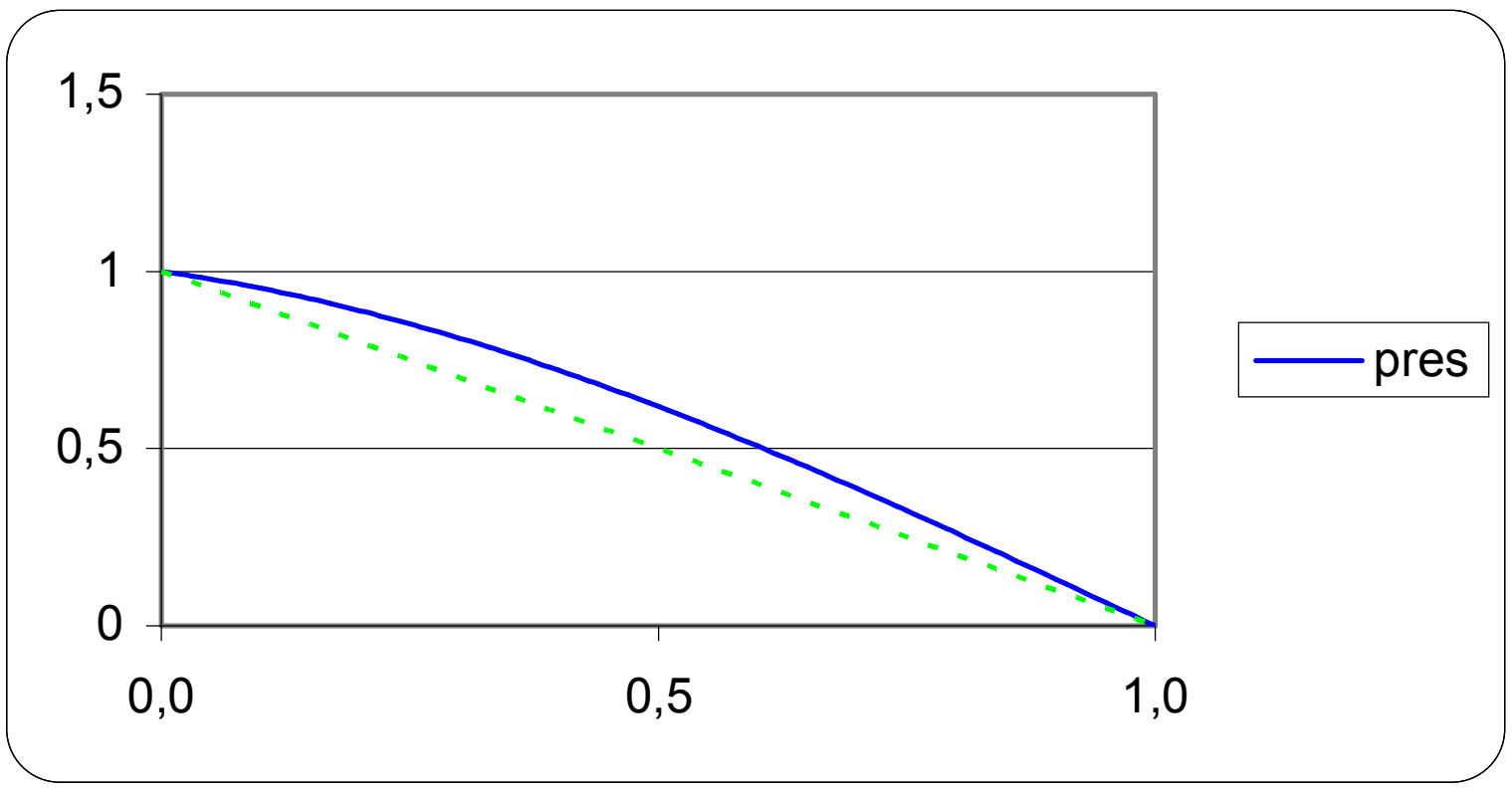

And even convexo-concave if there is a rotation

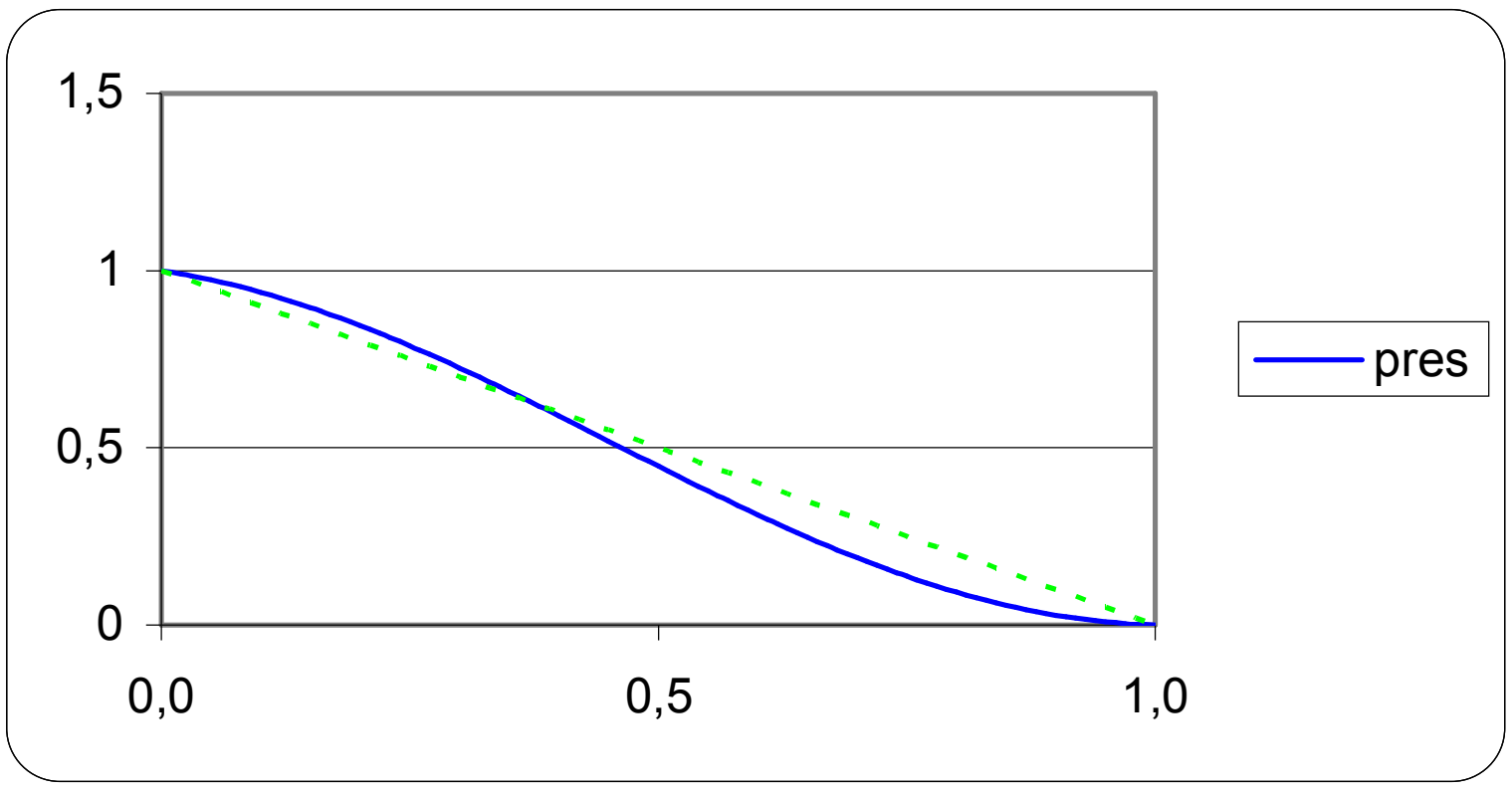

In summary we will notice that, apparently from the above-stated formulas, depending on temperature the stationary current of a viscous compressed liquid can be established in a cylindrical pipe in two ways: as with gradual increase in density on movement of a stream up to stopper formation, and with density reduction when speed of a stream increases up to failure of laminar. Last purely mathematical assumption is quite good for checking up by practical consideration.

[В заключение отметим, что, как видно из вышеприведённых формул, в зависимости от температуры стационарное течение вязкой сжимаемой жидкости может устанавливаться в цилиндрической трубе двумя способами: как с постепенным увеличением плотности по движению потока вплоть до образования пробки, так и с уменьшением плотности, когда скорость потока возрастает вплоть до срыва ламинарности. Последнее чисто математическое предположение неплохо бы проверить опытным путём.]

The literature